\documentstyle[preprint,revtex,eqsecnum]{aps}

\begin{document}
\draft
\widetext

%
%
%
\begin{title}
Towards a Canonical Formalism of Field Theory \\ on Discrete Spacetime
\end{title}
%
%
\author{Hiroshi YAMAMOTO,  Akihisa HAYASHI \\and\\ Takaaki HASHIMOTO}
\begin{instit}
Department of Applied Physics, Fukui University, 910, Fukui, Japan
\end{instit}
\author{Minoru HORIBE}
\begin{instit}
Department of Physics, Faculty of Education, Fukui University, 910 \\
Fukui, Japan
\end{instit}
%
%
%
%
\begin{abstract}
\baselineskip=0.75\baselineskip
   It is shown that the difficulties in formulating
the quantum field theory on discrete spacetime
appear already in classical dynamics of one
degree of freedom on discrete time.  The
difference equation of motion which maintains
a conserved quantity like energy has a
very restricted form that is not probably
derived by the least action principle.
On the other hand, the classical dynamics is possible
to be canonically formulated and quantized,
if the equation is derived from an action.
The difficulties come mainly from this
incompatibility of the conserved quantity and
the action principle.  We formulate a quantum field
theory canonically on discrete spacetime in
the case where the field equation is derived
from an action, though there may be no
exactly conserved quantity.  It may, however,
be expected that a conserved quantity exists
for a low "energy" region.
\end{abstract}
%
%
\pacs{PACS numbers: 03.70.+k, 11.10.-Z}
\section{{\bf Introduction}}
\label{sect1}
Since the classical dynamics was established by Newton, space and time have
been considered to be continuous.
In the theory the existence of space and time
is an ad hoc assumption and any physical object is supposed to be in this
spacetime. The location of the object is described by continuous coordinates
taken in this spacetime which satisfy a differential equation. The tradition
to describe physical phenomena in the continuous spacetime has been
inherited up to the contemporary physics and succeeded very much in almost
all area. However, when it was applied to
the field theory, it was confronted with the difficulty of infinity.
In the quantum electrodynamics the difficulty is
avoided by the renormalization
theory.  The theory is so beautiful and attractive that many people considered
the renormalizability as a fundamental principle of the field theory.
However, there are still some physicists who are
not totally satisfied with this theory. Dirac, for example, mentioned in a
lecture \cite{dirac} in 1968 as follows:

{\em "But still one is using working rules and not regular mathematics. Most
theoretical physicists nowadays appear to be satisfied with this situation,
but I am not. I believe that theoretical physics has gone on the wrong track
with such developments and one should not be complacent about it."} {\em
"Worrying over this point (=difficulty of infinity) may lead to an important
advance."}

Originally the spacetime in physics is an ad hoc assumption and its
continuity has never been confirmed by experiments.  Furthermore,
the existence of spacetime
is assumed independently of the existence of matter.
However, the spacetime is impossible to exist without matter.  How do we
measure a distance in space without matter?  The same question holds
for time.  The space is recognized by disposition of matter and
the time is recognized through the change of this disposition.

When we accept the above argument, we easily understand that spacetime
and matter are inseparable and can not exist independently.
In fact the existence of matter means at the same time the existence of
spacetime.  The particle physics of present-days seems to endeavor at
finding the fundamental or minimal constituents of matter.
If there exists the minimum constituent of the elementary particles,
there should exist the minimum element of spacetime. This means that
we have no means  to recognize a distance much smaller
than the length of this constituent. Considering in this way we may easily
accept the existence of minimum length in spacetime.
This length is probably in the same order as the size
of minimum constituents.  This is the reason, why we introduced the
discrete spacetime.

The experimental researches to examine the continuity of spacetime are rare.
Hawking and Ellis \cite{hawking} wrote as follows:

{\em "So far this continuity has been established for distance down to about
$10^{-15}$cm by experiments on pion scattering (Foley {\em et al.} (1967)).
It may be difficult to extend this to much smaller lengths as to do so
would require a particle of such high energy that several other particles
might be created and confuse the experiment. Thus it may be that a
manifold model for space-time is inappropriate for distances less than
$10^{-15}$cm and that we should use theories in which space-time has
some other structure on this scale."}

The above suggestion is along the same line as the lecture given by
Heisenberg at Cavendish Laboratory, Cambridge in 1949 \cite{heisenberg}.

{\em "Any future theory of the elementary particles must contain, besides
the fundamental constants $\hbar$ and $c$, a third fundamental constant
of the dimensions of a length or a mass.  This follows simply from the fact
that, on account of purely dimensional reasons, one cannot derive the
mass of an elementary particle from the constants $\hbar$ and $c$.
The actual mass of the main elementary particles suggests that this new
constant may be considered as a length $l$ of the order of magnitude
$l\sim10^{-13}$cm.  If the future theory contains such a constant in
whatever form, it is natural to assume that the usual correspondence
between the classical wave description and its quantum-theoretical
analogue only holds for very much greater than $l$, but fails in the
region of smaller distances."}

We have introduced the discrete spacetime and tried to formulate a canonical
quantum field theory on this
spacetime in the previous papers \cite{yamamoto1}.
There are also some approaches
in the same direction by other physicists \cite{yukawa}.
We have seen from the form of propagators
that there occurs no divergence difficulty at all.  In the case of fields
with interaction we have not yet succeeded to formulate the quantum theory
and to find a conserved quantity like energy that plays
a role of time developing generator. Therefore, the purpose of this paper is
to find the conserved quantity and the time developing operator and to
quantize the field canonically. Here we should emphasize our standpoint that
the field theory on discrete spacetime is never an approximation of
a continuous theory but  a true theory.  Conversely we consider that
the continuous theory is rather an approximation of the discrete theory.
At this point the theory on discrete spacetime is basically
different from the lattice gauge theory which is considered as an
approximation of the continuous theory.  While the lattice spacing in the
lattice gauge theory is expected to be set to zero at the end,
the fundamental length in the discrete spacetime is never set to zero
but remains finite.

In the theory on continuous spacetime Lagrangian is invariant under
infinitesimal translations and from Noether's theorem the corresponding
conserved quantities are easily obtained.  The energy and momenta
thus obtained are the translation generators of time and space
respectively.  In contrast to this,  Lagrangian on discrete spacetime
is invariant under the translation of a finite distance.  In this case
Noether's theorem is not applicable and we have no procedure to get
conserved quantities.  Moreover, we don't even know whether or not a
conserved quantity does exist.  In this sense Lagrangian might have no
important meaning.  To get a conserved quantity
and to find an operator of translation are different tasks.  The operator
of time translation is indispensable for canonical formulation. At the same
time conserved energy is necessary
for physical interpretation of the theory.
There is of course no verification that the generator and the energy
are identical to each other.

In the next section we show that the above mentioned is not only the
problem of field theory, but also the problem of classical dynamics
on discrete time.  There exists no conserved quantity, if time
is discretized simply in an equation of motion of classical dynamics.
We begin with an one-dimensional Newtonian equation and the time
derivatives are replaced by time differences in na{\"\i}ve way.
The Hamiltonian defined following to the Lagrangian formalism of
continuous time is not conserved.  The equation
is interpreted as a periodical map on the phase space of coordinate and
momentum.  If there exists a conserved quantity, the map remains on a
certain closed curve.  However, in our case the map behaves chaotically,
if the coordinate or momentum becomes large.
In order to get a conserved quantity the time should be carefully
discretized.  Furthermore, for the sake of physical interpretability
the area in the phase space should be conserved.
This gives another restriction
on the way of discretization.

Next we consider the case where the equation of motion is
derived from an action or Lagrangian.  The time translation operator is
possible to define using Poisson bracket.  However, it is still unknown
whether the conserved energy or the time translation generator exists and
is derived from the operator.  Only in special cases the conserved
quantity does exist and it is identical to the generator.  The advantage
of this method is the easiness of quantization, because it is enough to
replace the Poisson brackets by commutators.

In the section \ref{sect3} we show an attempt at formulating
a quantum field theory
with interaction on discrete spacetime.  In the free field case it is easily
seen that the time translation generator is identical to the conserved
energy operator.  In the case with special interaction the conserved
energy is obtained, but it is not the time translation generator
in general.  Conversely we can write down the time translation operator,
but we obtain neither generator nor energy in a compact form.  If we start
with the translation operator, we can quantize a scalar field canonically
and have a method to calculate S-matrix perturbatively.  In this sense
we have the quantum field theory on discrete spacetime, though we have
no conserved quantity.

In the section \ref{sect4} we show that our canonical
formalism is identical with path integral method on
discrete spacetime. The relation between
quantities on Euclidean and Minkowski spacetimes
is also discussed. Contrary to continuum theories,
they are not simply connected to each other by analytic
continuation.

The theory on discrete spacetime may seem a kind of cutoff theory and
hence the relativistic non-invariance comes to
question \cite{yamamoto2}.  At the end of
this introduction we cite again the words of Dirac \cite{dirac}:

{\em ''The relativistic invariance of the theory is then destroyed.
This is a pity, but it is a lesser evil than a departure from logic
would be.''}

%
%
\section{{\bf System with One Degree of Freedom on Discrete Time}}
\label{sect2}

   Though our final aim is to construct a canonical formalism
of field theories on discrete space-time, almost all
difficulties exist already in simple systems with few
degrees of freedom.  In this section, we will examine
the systems with only one degree of freedom on discrete time
and see what problems arise.

First, we will see that it is difficult to define Hamiltonian
by na{\"\i}ve discretization of the time in the usual canonical formalism.
When we want to discretize the time, there might be
many possible ways, but the simplest one is probably to
replace time derivatives by time differences.  If the minimum unit
of time is $\tau$, the difference operators are defined by

\begin{mathletters}
\begin{equation}
       \Delta^R x(t)\equiv \frac{1}{\tau}[x(t+\tau)-x(t)],
\end{equation}

\begin{equation}
       \Delta^L x(t)\equiv \frac{1}{\tau}[x(t)-x(t-\tau)].
\end{equation}
\end{mathletters}
Of course there are many operators other than
$\Delta^R$ and $\Delta^L$ that tend to the differential operator
$d/dt$ as $\tau\rightarrow 0$. However, we restrict ourselves to
consider only the above two for simplicity.

Now we assume the Lagrangian:
\begin{equation}
L(t)=\frac{1}{2}[\Delta^R x(t)]^2 -V[x(t)],
\end{equation}
where $V[x(t)]$ is potential and a function of $x(t)$.  Clearly the action:
\begin{equation}
S=\sum_{t} L(t) \label{action}
\end{equation}
is invariant under the finite translation:
\begin{equation}
t \longrightarrow t+\tau.
\end{equation}
The first problem is whether there exists a conserved quantity corresponding
to this invariance.

The equation of motion is obtained so that the action Eq.\ (\ref{action})
takes the least value.  That is
\begin{mathletters}
\begin{equation}
    \Delta^R \Delta^L x(t) = -\frac{\partial V[x(t)]}{\partial x(t)}
    \label{equation}
\end{equation}
or
\begin{equation}
  \frac{1}{\tau^2} \{x(t+\tau)-2x(t)+x(t-\tau)\}=-\frac{\partial
    V[x(t)]}{\partial x(t)}.
\end{equation}
\end{mathletters}
The momentum conjugate to $x(t)$ is
\begin{equation}
p(t)=\frac{\partial L(t)}{\partial \Delta^R x(t)}=
\Delta^R x(t)+\frac{\tau}{2}\frac{\partial V[x(t)]}{\partial x(t)}.
\end{equation}
The second term comes from the fact that $x(t)$ and $\Delta^R x(t)$
are not independent \cite{yamamoto1}:
\begin{equation}
x(t) = \frac{1}{2}\{x(t+\tau)+x(t)\}-\frac{1}{2}\{x(t+\tau)-x(t)\}.
\end{equation}
Hence we have
\begin{equation}
\frac{\partial x(t)}{\partial \Delta^R x(t)}=-\frac{\tau}{2}.
\end{equation}
According to the usual Lagrangian formalism the conserved quantity is
Hamiltonian defined by
\[
H  =  p(t)\Delta^R x(t)-L(t)
\]
\begin{equation}
\  =  \frac{1}{2}[\Delta^Rx(t)]^2 +V[x(t)]+\frac{\tau}{2}
\frac{\partial V[x(t)]}{\partial x(t)}\Delta^Rx(t).
                   \label{hamiltonian}
\end{equation}

Let us see what happens if the potential is given by $V[x(t)]=
\Lambda x^n(t)/n$.  The equation of motion is
\[
\Delta^R \Delta^L x(t) =-\Lambda x^{n-1}(t),
\]
and Hamiltonian defined by Eq.\ (\ref{hamiltonian}) is given by
\begin{equation}
H=\frac{1}{2}[\Delta^R x(t)]^2+\frac{1}{n}\Lambda x^n(t)+\frac{1}{2}
\Lambda x^{n-1}(t)\{x(t+\tau)-x(t)\}.
\end{equation}
It is readily verified that the above Hamiltonian
is conserved only when $n=1$ (uniform gravitation)
or $n=2$ (harmonic oscillator).
Thus we conclude that the Hamiltonian
Eq.\ (\ref{hamiltonian}) obtained by the na{\"\i}ve discretization
is not conserved in general:
\begin{equation}
\Delta^L H\neq 0.
\end{equation}
Furthermore, the above Hamiltonian cannot be
regarded as a time-developing generator.

An interesting question would be whether there exists
a conserved quantity corresponding to energy in the
case of nonlinear forces, and if it is the case, how
it is related to the time-developing generator.  We have
two different approaches to study these problems. The first
one is described in Subsec.\ \ref{subsect_2a}, where we
show that the equation of motion can be discretized so
that the system has an energy-like conserved quantity,
though it is not generally the time-developing generator.
In the second approach given in Subsec.\ \ref{subsect_2b},
we can define a time-developing operator and construct
a canonical formalism. Although the construction is rather
formal, this method is useful if the time scale of the
motion is much larger than the minimum unit of time $\tau$.
Unfortunately the two approaches are not related to each
other except for only special cases.

\subsection{{\bf Conserved quantity}}
\label{subsect_2a}
   We assume the equation of motion:
\begin{equation}
\Delta^R \Delta^L x(t) = f(t). \label{equationf}
\end{equation}
When the equation is derived by the principle of least action, it
 is the same as Eq.\ (\ref{equation}).
We can easily verify that the following quantity:
\begin{equation}
\tilde{H} = \frac{1}{2}[\Delta^Rx(t)]^2 -
               \frac{1}{2}\sum_{t^{\prime} =t_0}^{t}\{x(t^{\prime}+\tau)
               -x(t^{\prime}-\tau)\}f(t^{\prime}) \label{htilde}
\end{equation}
is conserved:
\begin{equation}
\Delta^L\tilde{H} = 0.
\end{equation}
It must be noticed here that $\tilde{H}$ includes the summation of
time-dependent quantities over all times from a certain past $t_0$ to
the present $t$.

 When the force is given by $f(t)=-\Lambda x^{n-1}(t)$ as before,
$\tilde H$ takes the form:

\[
\tilde{H}=\frac{1}{2}[\Delta^Rx(t)]^2+\frac{1}{2}\Lambda \sum_{t^{\prime}
=t_0}^{t}\{x(t^{\prime}+\tau)-x(t^{\prime}-\tau)\}x^{n-1}(t^{\prime}).
\]

If $n=1$ or $n=2$, the terms in the summation cancel each other
and $\tilde H$ is reduced to $H$ obtained by the na{\"\i}ve discretization.
We, however, see that the terms in summation (\ref{htilde}) do not
cancel each other in general.  Because the conserved quantity should be
expressed by $x(t)$ and $x(t+\tau)$ or equivalently $x(t)$ and $p(t)$:
\begin{equation}
\tilde{H}=\tilde{H}[x(t),x(t+\tau)] \ \ \ {\rm or} \ \ \ \tilde{H}[x(t),p(t)],
\label{h_condition}
\end{equation}
$\tilde{H}$  cannot generally be called a conserved quantity.

So far, we discretized only the term of time derivative in the equation
of motion on continuous time.  However, if we consider our guiding
principle that the discretized equation should return to the original
differential equation of motion as $\tau\rightarrow 0$, we may not disregard
the discretization of the interaction term.  That is, if the force term
$f(t)$ satisfies
\begin{equation}
\lim_{\tau\rightarrow 0}f(t)=-\frac{\partial V[x(t)]}{\partial x(t)},
\label{f_condition}
\end{equation}
the difference equation (\ref{equationf}) tends
to the original differential equation:
\begin{equation}
\frac{d^2x}{dt^2}=-\frac{\partial V}{\partial x}.
\end{equation}
Therefore, we may discretize the interaction term so that the obtained
equation has a conserved quantity.

First we suppose the following force:
\begin{equation}
f(t)=-\frac{V(t+\tau)-V(t-\tau)}{x(t+\tau)-x(t-\tau)}
 \label{force}
\end{equation}
where we abbreviated $V[x(t)]$ to $V(t)$ for simplicity.  The force
satisfies obviously the above condition (\ref{f_condition}).
However, the equation of motion:
\begin{equation}
\Delta^R\Delta^Lx(t)=-\frac{V(t+\tau)-V(t-\tau)}{x(t+\tau)-x(t-\tau)},
\end{equation}
does not seem to be
derived from Lagrangian by the principle of least action as usual.
Now we substitute Eq.\ (\ref{force})
in Eq.\ (\ref{htilde}) and we have
\begin{equation}
\tilde{H}=\frac{1}{2}[\Delta^Rx(t)]^2+\frac{1}{2}\{V(t+\tau)+V(t)\}
-\frac{1}{2}\{V(t_0)+V(t_0-\tau)\}.
\end{equation}
Since $\tilde{H}$ includes only $x(t)$ and $x(t+\tau)$, or satisfies
the condition Eq.\ (\ref{h_condition}), we may call it the conserved quantity.

There exist other possibilities.  If the potential is
\begin{equation}
V(t)=\frac{\Lambda}{2m}\,x^{2m}(t), \hspace{1cm} (m:\mbox{integer})
\end{equation}
we assume
\begin{equation}
f(t)=-\frac{\Lambda}{m}\cdot\frac{x^m(t+\tau)-x^m(t-\tau)}
{x(t+\tau)-x(t-\tau)}
x^m(t),
\end{equation}
and then we have
\begin{equation}
\tilde{H}=\frac{1}{2}[\Delta^Rx(t)]^2+\frac{\Lambda}{2m}x^m(t+\tau)x^m(t)
-\frac{\Lambda}{2m}x^m(t_0)x^m(t_0-\tau).
\end{equation}
In the case where the potential is
\begin{equation}
V(t)=\frac{\Lambda}{2m+1}x^{2m+1}(t),
\end{equation}
substituting
\begin{eqnarray}
f(t)  =-\frac{\Lambda}{2m+1}\biggl\{ && \frac{x^m(t+\tau)-x^m(t-\tau)}
{x(t+\tau)-x(t-\tau)}x^{m+1}(t) \nonumber \\
&& -\frac{x^{m+1}(t+\tau)-x^{m+1}(t-\tau)}
{x(t+\tau)-x(t-\tau)}x^m(t) \biggr\},
\end{eqnarray}
we have
\begin{eqnarray}
\tilde{H}  = \frac{1}{2}[\Delta^Rx(t)]^2+ && \frac{\Lambda}{2(2m+1)}
x^m(t+\tau)x^m(t)\{x(t+\tau)+x(t)\}  \nonumber \\
&& -\frac{\Lambda}{2(2m+1)}x^m(t_0)x^m(t_0-\tau)\{x(t_0)
+x(t_0-\tau)\}.
\end{eqnarray}

    Thus we found that for an arbitrary potential the equation
of motion can be discretized so that the system still has
an energy-like conserved quantity. As we will see in section\ref{sect3},
the similar discretization method can be applied to field theories.

\subsection{{\bf Time developing operator}}
\label{subsect_2b}

    The approach mentioned in Subsec.\ \ref{subsect_2a} does
not generally give the canonical conjugate momentum which,
together with the coordinate $x$, defines an area preserving
dynamics. Furthermore, it is hard to establish relation between
the conserved quantity obtained in this way and the time
developing generator of the system. The problems will be more
serious, when we try to quantize the system canonically.  We
speculate that the difficulties arise from the fact that the
newly discretized equation of motion can not generally be derived
from the least action principle. Keeping this in mind, we next
attempt to define the time developing operator in the case where
the equation of motion is derived from the action (\ref{action}).
Once the time developing generator is defined, it is clearly a
conserved quantity.

    We begin with the equation of motion Eq.\ (\ref{equation}),
which is derived from the na{\"\i}vely discretized action
Eq.\ (\ref{action}). Defining the canonical momentum $p(t)$
conjugate to $x(t)$ as
\begin{equation}
p(t)=\Delta^R x(t),
\label{reasonable_p}
\end{equation}
we look for the
conserved quantity in a power series of $\tau$:
\begin{equation}
H_{\tau} =\frac{1}{2}p^2(t)+V[x(t)]+\sum_{n=1}^{\infty}\tau^n H_n
[x(t),p(t)]. \label{tau_expansion}
\end{equation}
Of course $H_{\tau}$ should tend to the Hamiltonian of continuous time
for $\tau\rightarrow 0$.

Now we consider the time developing operator of the form :
\begin{equation}
\Omega (t+\tau)= U^{-1}\Omega (t),
 \label{udef}
\end{equation}
\begin{equation}
U^{-1}\equiv e^{-\tau \{\frac{1}{2}p^2(t),\hspace{0.6cm}\}_{\rm P.B.}}
e^{-\tau \{V[x(t)],\hspace{0.6cm}\}_{\rm P.B.}},
\end{equation}
where
\begin{equation}
e^{-\tau \{P,\hspace{0.6cm}\}_{\rm P.B.}} \Omega = \sum_{n=0}^{\infty}
\frac{(-\tau)^n}{n!}
    \{P,\{P,\{\cdots\{P,\Omega \overbrace {\}\cdots\}\}\}_{\rm P.B.}}^{n}
\end{equation}
and $\{P,Q\}_{\rm P.B.}$ means Poisson bracket:
\begin{equation}
\{P,Q\}_{\rm P.B.} =
\frac{\partial P}{\partial x}\frac{\partial Q}{\partial p}-
\frac{\partial Q}{\partial x}\frac{\partial P}{\partial p}.
 \label{poisson}
\end{equation}
It should be noted that we defined $U^{-1}$ rather than
$U$ as in Eq.\ (\ref{udef}) so that $U$ will be the usual
time developing operator after quantization.
We see easily
\begin{mathletters}
\begin{equation}
  x(t+\tau)=U^{-1}x(t)=x(t)+\tau p(t), \label{eqmap}
\end{equation}
\begin{equation}
p(t+\tau)=U^{-1}p(t)=p(t)-\tau\frac{\partial V[x(t+\tau)]}
{\partial x(t+\tau)}.
\end{equation}
\end{mathletters}
It is clear that the above equations reproduce
the original equation of motion Eq.\ (\ref{equation}).

The question is whether the operator $U$ can be expressed in the form:
\begin{equation}
U^{-1}=e^{-\tau \{H_{\tau},\hspace{0.6cm}\}_{\rm P.B.}}.
\end{equation}
If it is possible, $H_{\tau}$ is a conserved quantity, since
\begin{equation}
H_{\tau}(x(t+\tau),p(t+\tau))=U^{-1}H_{\tau}(x(t),p(t))=H_{\tau}(x(t),p(t)).
\end{equation}
Using Hausdorf's formula:
\begin{equation}
e^{\tau X}e^{\tau Y} = e^{\tau (X+Y)+(\tau^2/2)[X,Y]+\cdots},
\end{equation}
we formally obtain
\begin{equation}
H_{\tau}=\frac{1}{2}p^2(t)+V[x(t)]+\frac{\tau}{2}p(t)V^{\prime}[x(t)]
+\frac{\tau^2}{12}\{p^2(t)V^{\prime \prime}[x(t)]+V^{\prime 2}[x(t)]\}
+O(\tau^3). \label{htau_exp}
\end{equation}
In fact this quantity is of the form of $\tau$-expansion
in Eq.\ (\ref{tau_expansion}).
 The same approach was also studied by Yoshida \cite{yoshida}
in the context of numerical integration method for continuous
equations of motion.

   In the case of the uniform gravitational potential,
the power series in $\tau$ ends at the second order as follows:
\begin{eqnarray*}
H_{\tau} & = & \frac{1}{2}p^2(t)+gx(t)+\frac{\tau}{2}gp(t)+\frac{\tau^2}
               {12}g^2 \\
         & = & \frac{1}{2}[\Delta^Rx(t)]^2+\frac{1}{2}g\{x(t+\tau)+x(t)\}
               +\frac{\tau^2}{12}g^2.
\end{eqnarray*}
This agrees with $H$ and $\tilde H$ obtained previously,
except for a constant term.

    When the potential is harmonic, this series in $H_{\tau}$
again takes the closed form:
\begin{eqnarray*}
H_{\tau} & = & K(\tau)\{p^2(t)+\omega^2 x^2(t)+\tau \omega^2p(t)x(t)\} \\
         & = & K(\tau)\{[\Delta^Rx(t)]^2+\omega^2x(t)x(t+\tau)\},
\end{eqnarray*}
with
\[
K(\tau)= \frac{1}{\omega\tau\sqrt{4-(\omega\tau)^2}}
\cos^{-1}(1-\frac{(\omega\tau)^2}{2}).
\]
This also agrees with the conserved quantity obtained before,
except for a factor.  In passing
we see
\[
\lim_{\tau\rightarrow 0}K(\tau)=\frac{1}{2},
\]
thus $H_{\tau}$ recovers the harmonic hamiltonian in the continuous
time as $\tau$ goes to 0.

    In the case of nonlinear potential we cannot have any explicit form of
$H_{\tau}$.  However, it does not mean non-convergence of the series.
As an example we study numerically the nonlinear iterative map
Eqs.\ (\ref{eqmap},b) with the potential $ V(x)=\Lambda x^4/4 $.
In Fig.\ 1 we plot points  generated by the map in terms of
scaled variables $\tilde x=\tau\sqrt{\Lambda}\,x$ and $\tilde
p=\tau^2\sqrt{\Lambda}\,p$. For initial values, we fix $\tilde p(0)$
to be 0 and take several different $\tilde x(0)$'s.
We see that for relatively small initial values of $\tilde x(0)$
the map draws a smooth curve, KAM-curve, whereas the map shows
chaotic behaviour for larger initial values of $\tilde x(0)$.
 The smooth curve should be defined by a function of $x(t)$ and
$p(t)$.  This means there exists a conserved quantity, which
is presumably
$H_{\tau}$. In conclusion we may consider that the expansion
Eq.\ (\ref{htau_exp}) converges for small $\tilde x(t)$ and $\tilde p(t)$.

    As we have seen in Subsec.\ \ref{subsect_2a},
we can discretize the equation of motion so that the
system has a conserved quantity, but in this case
it is not easy to define the canonical variables and
construct the time developing operator. On the other
hand, in the second approach (Subsec.\ \ref{subsect_2b}),
one can construct the time developing operator in terms
of canonical variables and obtain the formally conserved
Hamiltonian $H_{\tau}$. Although we cannot say much about
the convergence of $H_{\tau}$, we speculate from the
example that the series converge, if the minimum unit
of time $\tau$ is much less than the time scale of
the motion considered. Assuming the convergence, we
proceed to quantize the dynamics in the second approach.
The quantization is straightforward: the canonical
variables $x$ and $p$ are replaced by the operators
$\hat x$ and $\hat p$ satisfying the usual canonical
commutation relation $[\,\hat x,\hat p\,]=i$.
Using the quantized time-developing operator
\begin{equation}
  \hat U=e^{-i\tau V(\hat x)}e^{-i\tau\hat p^2/2}=
         e^{-i\tau H_\tau(\hat x,\hat p)},
\end{equation}
we easily obtain
\begin{mathletters}
\begin{equation}
  \hat x(t+\tau)=\hat U^{-1}\hat x(t)\hat U = \hat x(t)+\tau \hat p(t),
\end{equation}
\begin{equation}
 \hat p(t+\tau)=\hat U^{-1}\hat p(t)\hat U = \hat p(t)-\tau
        \frac{\partial V[\hat x(t+\tau)]}{\partial \hat x(t+\tau)}.
\end{equation}
\end{mathletters}
which have the same form as the classical equations of motion.
In the Schr\"odinger representation,
the wave function $\psi_S\,(t)$ develops
in time according to the following integrated
form of the Schr\"odinger equation:
\begin{equation}
 \psi_S\,(t+\tau) = \hat U \psi_S\,(t).
\end{equation}
Since the equation of motion is derived from the action
in this case, we could quantize the system by the use
of the path integral method. In the last section \ref{sect4},
we will discuss how our canonical quantization is related
to the path integral.

\section{{\bf Field Theory}}
\label{sect3}
%
	In this section we consider the formulation of field theory on discrete
spacetime. The field theory is not much different from the classical dynamics
except for that it is a multi-freedom system, because every point of discrete
spacetime has a dynamical freedom.  Hence we refer to the foregoing section
and consider a scalar field.

We begin with the following difference equation, which is obtained by
replacing Klein-Gordon operator with a na\"{\i}ve difference operator:
\begin{equation}
\sum^{3}_{\mu,\nu=0}{\eta}^{\mu\nu}\Delta^{R}_{\mu}
\Delta^{L}_{\nu}\phi
\left(
x\right)=G\left(\phi\left(x\right)\right),
\label{threeone}
\end{equation}
where
\begin{mathletters}
\begin{equation}
\Delta^{R}_{\mu}\phi\left(x\right)={1 \over \sigma\left(\mu\right)}
\left[\phi\left(x+\hat{\sigma}\left(\mu\right)\right)-\phi
\left(x\right)\right],
\label{threetwoa}
\end{equation}
\begin{equation}
\Delta^{L}_{\mu}\phi\left(x\right)={1 \over \sigma\left(\mu\right)}
\left[\phi\left(x\right)-\phi\left(x
-\hat{\sigma}\left(\mu\right)\right)\right].
\label{threetwob}
\end{equation}
\end{mathletters}
	$\hat{\sigma}\left(\mu\right)$ is a vector directing to $\mu$-axis
with length $\sigma\left(\mu\right)$.  We assume for simplicity
\begin{equation}
\sigma\left(0\right)=\tau\;\;\;,\;\;\;\sigma\left(i\right)=\sigma\;\;\;\;\;
\left(i=1,2,3,\right).
\label{threethree}
\end{equation}
The spacetime points are then expressed by
\begin{equation}
x=\sum^{3}_{\mu=0}\hat{\sigma}\left(\mu\right)n_{\mu}\;\;\;\;\;\;
\left(n_{\mu}\;\;:\;\;\rm{integers}\right).
\label{threefour}
\end{equation}
We are considering Minkowski spacetime and the metric tensor is
$\eta^{\mu\nu}=\left(1,-1,-1,-1\right)$.
Equation~(\ref{threeone}) is then written as
\begin{eqnarray}
{1 \over \tau^2}& &\left[\phi\left(x^0+\tau,{\bf x} \right)-2\phi\left(
x^0,{\bf x}\right)+\phi\left(x^0-\tau,{\bf x}\right)\right]\nonumber\\
& &={1 \over {\sigma}^2}\sum^{3}_{i=1}\left[\phi\left(x^0,
{\bf x}+\hat{\sigma}\left(i \right)\right)-2\phi\left(x^0,{\bf x} \right)+
\phi \left(x^0,{\bf x}-\hat{\sigma}\left(i \right)\right) \right]
+G\left(\phi\left(x\right)\right).
\label{threefive}
\end{eqnarray}
$G\left(\phi\right)$ represents the mass term and
interaction terms.  In the continuous spacetime limit or in the continuous
approximation $G\left(\phi\right)$ agrees with what is
derived from the interaction Lagrangian or potential
$V\left(\phi\right)$:
\begin{equation}
\lim_{\sigma,\tau \rightarrow 0} G(\phi)=-{\partial V \over
\partial \phi}\;\;.\label{threesix}
\end{equation}
\subsection{{\bf Free field and canonical quantization}}\label{free}
	In the field theory on continuous spacetime there are two ways of
quantization, that is, path integral method and canonical quantization.  If
one wants to quantize a field by path integral, it is necessary to have the
Lagrangian or action. Equation~(\ref{threeone}) or (\ref{threefive})
is not always derived from Lagrangian, because $G\left(\phi\right)$ is not
necessarily derived from a potential $V\left(\phi\right)$:
\[G(\phi) \not=-{\partial V \over \partial \phi}\;\;.\]
The case where $G\left(\phi\right)=-\partial{V} / \partial{\phi}$ will be
considered in Subsec.\ \ref{perturb}.

Now we assume
\begin{equation}
G\left(\phi\right)=-m^2\phi,
\label{threeseven}
\end{equation}
that corresponds to a free field with mass $m$. The field equation is
\begin{equation}
\sum^{3}_{\mu,\nu=0}{\eta}^{\mu\nu}\Delta^{R}_{\mu}
\Delta^{L}_{\nu}\phi
\left(
x\right)+m^2\phi\left(x\right)=0.
\label{threeeight}
\end{equation}
This equation is identical with the following canonical equation:
\begin{mathletters}
\begin{equation}
\phi\left(x^0+\tau,{\bf x}\right)=\phi\left(x^0,{\bf x}\right)+\tau
\pi\left(x^0,{\bf x}\right),
\label{threeninea}
\end{equation}
\begin{equation}
\pi\left(x^0+\tau,{\bf x}\right)=\pi\left(x^0,{\bf x}\right)
-\tau\left[m^2\phi\left(x^0+\tau,{\bf x}\right)-\sum^{3}_{i=1}\Delta^R_i
\Delta^L_i\phi\left(x^0+\tau,{\bf x}\right)\right].
\label{threenineb}
\end{equation}
\end{mathletters}
If we regard $\pi\left(x^0,{\bf x}\right)$ as the conjugate to $\phi
\left(x^0,{\bf x}\right)$, Poisson bracket is
\begin{equation}
\left\{\phi\left(x^0,{\bf x}\right),\pi\left(x^0,{\bf x'}\right)
\right\}_{\rm P.B.}
=\delta_{{\bf x},{\bf x'}}.
\label{threeten}
\end{equation}
Using Equations~(\ref{threeninea}) and~(\ref{threenineb}), we easily verify
\begin{eqnarray}
\lefteqn{\left\{\phi\left(x^0+\tau,{\bf x}\right),
\pi\left(x^0+\tau,{\bf x'}\right)\right\}_{\rm P.B.}}
\;\;\;\;\;\;\;\;\;\;\nonumber\\
& = & \sum_{{\bf y}}\left[{
      \partial\phi\left(x^0+\tau,{\bf x }\right) \over
      \partial\phi\left(x^0,     {\bf y }\right)}{
      \partial\pi \left(x^0+\tau,{\bf x'}\right)  \over
      \partial\pi \left(x^0,     {\bf y }\right)}-{
      \partial\phi\left(x^0+\tau,{\bf x }\right) \over
      \partial\pi \left(x^0,     {\bf y }\right)}{
      \partial\pi \left(x^0+\tau,{\bf x'}\right)  \over
      \partial\phi\left(x^0,     {\bf y }\right)}\right]\nonumber\\
& = & \delta_{{\bf x},{\bf x'}}.
\label{threeeleven}
\end{eqnarray}
This means that Poisson bracket is conserved.
The equation~(\ref{threeeight}) is easily solved to give
\begin{equation}
\phi\left(x\right)=\left({\sigma \over 2\pi}
                   \right)^{3 \over 2}
                   \int_{R}d^3k\sqrt{{\tau \over 2\sin{\tau\omega}
                                     }
                                    }
\left[a\left({\bf k}
       \right)e^{-i\left({\omega}x^0-{\bf k}\cdot{{\bf x}}
                   \right)
                }+ c.c.
\right],
\label{threetwelve}
\end{equation}
where $\omega$ is defined by
\begin{equation}
{1 \over \tau^2}{\sin^2{{\tau\omega \over 2}}}=
{1 \over \sigma^2}\sum^{3}_{i=1}\sin^2{{{\sigma}k_i \over 2}}
+ {m^2 \over 4},
\label{threethirteen}
\end{equation}
and $a\left({\bf k}\right)$ is an arbitrary function of ${\bf k}$.
The domain of integration $R$ is
\begin{equation}
-{\pi \over \sigma} \leq k_{i} \leq {\pi \over \sigma}\;\;\;\;\;\;
\left(i=1,2,3\right).
\label{threefourteen}
\end{equation}
Using Eq.~(\ref{threeninea}), $\pi\left(x\right)$
is also expressed by
$a\left({\bf k}\right)$ and $a^{\ast}\left({\bf k}\right)$.  From
Eq.~(\ref{threeten}) we have Poisson bracket of
$a\left({\bf k}\right)$ and $a^{\ast}\left({\bf k}\right)$:
\begin{equation}
\left\{a\left({\bf k}\right),a^{\ast}\left({\bf k'}\right)\right\}
_{\rm P.B.}=-i\delta\left({\bf k}-{\bf k'}\right).
\label{threefifteen}
\end{equation}
     Now we look for the time developing generator $H_{0}$, which satisfies
\begin{equation}
\phi\left(x^0+\tau,{\bf x}\right)=
e^{-\tau\left\{H_{0},\;\;\;\;\;\right\}_{\rm P.B.}}
\phi\left(x^0,{\bf x}\right).
\label{threesixteen}
\end{equation}
The notation of the above equation is the same as Eq.~(\ref{poisson}).
  This means
\begin{mathletters}
\begin{equation}
e^{-\tau\left\{H_{0},\;\;\;\;\;\right\}_{\rm P.B.}}
a\left({\bf k}\right)=e^{-i\omega\tau}
a\left({\bf k}\right),
\label{threeseventeena}
\end{equation}
\begin{equation}
e^{-\tau\left\{H_{0},\;\;\;\;\;\right\}_{\rm P.B.}}
a^{\ast}\left({\bf k}\right)=e^{i\omega\tau}
a^{\ast}\left({\bf k}\right).
\label{threeseventeenb}
\end{equation}
\end{mathletters}
Therefore, we have
\begin{mathletters}
\begin{equation}
H_{0}={1 \over 2}\int_{R}d^3k\;\;\;\omega\left[
a       \left({\bf k}\right)a^{\ast}\left({\bf k}\right)+
a^{\ast}\left({\bf k}\right)a       \left({\bf k}\right)\right],
\label{threeeighteen}
\end{equation}
or using $\phi\left(x\right)$ and $\pi\left(x\right)$,
\begin{eqnarray}
H_{0} & = & \left({\sigma \over 2\pi}\right)^3\sum_{{\bf x},{\bf x'}}
            \int_{R}d^3k \;\cos{{\bf k}\cdot\left({\bf x}-{\bf x'}\right)}
\nonumber\\
          & & \times \biggr[{\omega\tau \over 4
              \tan{\omega\tau/2}}
              \pi\left(x^0,{\bf x}\right)\pi\left(x^0,{\bf x'}\right)+
               {\sin{\omega\tau} \over 2\omega\tau}
              \omega^2\phi\left(x^0,{\bf x}\right)
              \phi\left(x^0,{\bf x'}\right)
\nonumber\\
        & &\;\;\;\;\;- {\omega\sin{\omega\tau} \over 4}
             \left\{\phi\left(x^0,{\bf x }\right)\pi \left(x^0,{\bf x'}\right)
              +     \pi \left(x^0,{\bf x'}\right)\phi\left(x^0,{\bf x }\right)
             \right\}
                     \biggr].
\label{threenineteen}
\end{eqnarray}
\end{mathletters}
This $H^0$ is rather complicated, but the time developing operator $U_{0}$ is
expressed in a more simplified form:
\begin{eqnarray}
U_{0}^{-1} & = &e^{-\tau\left\{H_{0}\;,\;\;\;\;\;\right\}_{\rm P.B.}
             }\nonumber\\
      & = &  \exp{\left\{-\tau\sum_{{\bf x}}{1 \over 2}
                     \pi^2\left(x^0,{\bf x}\right)\;\;,\;\;\;\;\;
                  \right\}_{\rm P.B.}
                 }\nonumber\\
&   & \times \exp{\left\{-\tau\sum_{{\bf x}}
                   \left[{1 \over 2}\sum^{3}_{i=1}
                      \left(\Delta^{R}_{i}\phi\left(x^0,{\bf x}\right)
                      \right)^2+{1 \over 2}m^2\phi^2\left(x^0,{\bf x}\right)
                   \right]\;\;,
                   \;\;\;\;\;
                \right\}_{\rm P.B.}
               }\;\;\;.
\label{threenineteendash}
\end{eqnarray}

	As we have seen above, the conserved energy and the time developing
generator are identical in the case of free field.  Therefore, the field is
easily quantized by replacing Poisson bracket with commutation relations, e.g.
\begin{equation}
\left\{A,B\right\}_{\rm P.B.}\;\;\;\longrightarrow\;\;\;
{1 \over i}\left[A,B\right]
={1 \over i}\left(AB-BA\right).
\label{threetwenty}
\end{equation}
The time developing operator is
\begin{eqnarray}
U_{0}\left(\phi,\pi\right) & = & e^{-i{\tau}H_{0}}\nonumber\\
& = & \exp{\left[-i\tau\sum_{\bf x}
              \left\{{1 \over 2}\sum^{3}_{i=1}
                 \left(\Delta^{R}_{i}\phi\right)^2+{1 \over 2}m^2\phi^2
              \right\}
           \right]
          }
      \exp{\left[-i\tau\sum_{\bf x}
              {1 \over 2}\pi^2
           \right]
          },
\label{threetwentyone}
\end{eqnarray}
with
\begin{equation}
\phi\left(x^0+\tau,{\bf x}\right)=U_{0}^{-1}\left(\phi,\pi\right)
\phi\left(x^0,{\bf x}\right)      U_{0}     \left(\phi,\pi\right).
\label{threetwentytwo}
\end{equation}
Feynman propagator of the fields ~\cite{yamamoto1} is
\begin{eqnarray}
\lefteqn{\langle0\mid T\left(
\phi\left(x\right)\phi\left(x'\right)
\right)\mid0\rangle =
\left({\sigma \over 2\pi}\right)^3\int_{R}d^3k
\left({\tau \over 2\sin{\omega\tau}}\right)}\;\;\;\;\;\;\;\;\;\;\;
\nonumber\\
& &\times\biggr[\theta\left(x^0-x'^0\right)
   \exp{\left\{-i\omega\left(x^0-x'^0\right)
          +i{\bf k}\cdot\left({\bf x}-{\bf x'}\right)
        \right\}
       }\nonumber\\
& &\;\;\;\;\;+\theta\left(x'^0-x^0\right)
   \exp{\left\{i\omega\left(x^0-x'^0\right)
          -i{\bf k}\cdot\left({\bf x}-{\bf x'}\right)
        \right\}
       }\biggr]\nonumber\\
& = & i
\left(\prod^{3}_{\mu=0}
     \int^{{  \pi \over \sigma\left(\mu\right)}}
          _{-{\pi \over \sigma\left(\mu\right)}}
     {\sigma\left(\mu\right) \over 2\pi}dk_\mu
\right)
\exp{\left\{-ik_\nu\left(x^\nu-x'^\nu\right)\right\}}\nonumber\\
& &\times\left[\sum^{3}_{\mu,\nu=0}\eta^{\mu\nu}
                 {4 \over \sigma\left(\mu\right)\sigma\left(\nu\right)
                 }
             \sin{{\sigma\left(\mu\right)k_{\mu}} \over 2}
             \sin{{\sigma\left(\nu\right)k_{\nu}} \over 2}
             -m^2+i\epsilon
         \right]^{-1}
    .
\label{threetwentythree}
\end{eqnarray}
\subsection{{\bf Interacting field with a conserved quantity}}\label{int.con}
	In this subsection we consider a field with interaction , where we
regard the existence of a conserved quantity as the most important requirement
.  Under the condition the form of interaction $G\left(\phi\right)$ in
Eq.~(\ref{threeone}) or ~(\ref{threefive}) is very much restricted.
We show two cases in the following.

\noindent
	a) We assume Eq.~(\ref{threeone}) or ~(\ref{threefive}) and
\begin{equation}
G\left(\phi\right)=-{g \over 2}\left\{
\phi\left(x^0+\tau,{\bf x}\right)+\phi\left(x^0-\tau,{\bf x}\right)\right\}
\phi^2\left(x\right)-m^2\phi\left(x\right),
\label{threetwentyfour}
\end{equation}
that tends to
\begin{equation}
\lim_{\sigma,\tau \rightarrow 0} G(\phi)=-{\partial V \over
\partial \phi}\;\;\;\;\;\;\;\rm{with}\;\;\;\;\;\;\;V={g \over 4}\phi^4+
{m^2 \over 2}\phi^2,
\label{theetwentyfive}
\end{equation}
in the continuous limit. In this case the field equation~(\ref{threeone}) or
{}~(\ref{threefive}) is rewritten canonically and shown to have a conserved
quantity corresponding to the energy.  It is easily understood if we notice
that the Equation is a simple extension of the classical dynamics with $x^4$
-potential shown in section II.
	The equation is written in a canonical form:
\begin{mathletters}
\begin{equation}
\phi\left(x^0+\tau,{\bf x}\right)=\phi\left(x^0,{\bf x}\right)+\tau
\pi\left(x^0,{\bf x}\right),
\label{threetwentysixa}
\end{equation}
\begin{equation}
\pi\left(x^0+\tau,{\bf x}\right)=\pi\left(x^0,{\bf x}\right)
+\tau\sum_{{\bf x'}}{\bf F}\left(\phi\left(x^0+\tau,{\bf x}\right)
\right)_{{\bf x},{\bf x'}}\phi\left(x^0+\tau,{\bf x'}\right),
\label{threetwentysixb}
\end{equation}
\end{mathletters}
where
\begin{equation}
{\bf F}\left(\phi\left(x^0+\tau,{\bf x}\right)\right)_{{\bf x},{\bf x'}}
=\left\{{1 \over 2}g\phi^2\left(x^0,{\rm x}\right)+{1 \over \tau^2}
 \right\}^{-1}
 \left({2 \over \tau^2}\delta_{{\bf x},{\bf x'}}
       -{\bf \Omega}_{{\bf x},{\bf x'}}
 \right)
-{2 \over \tau^2}\delta_{{\bf x},{\bf x'}},
\label{threetwentyseven}
\end{equation}
and
\begin{equation}
{\bf \Omega}_{{\bf x},{\bf x'}}=m^2\delta_{{\bf x},{\bf x'}}
-{1 \over \sigma^2}\sum^{3}_{i=1}
\left(\delta_{{\bf x}+\hat{\sigma}\left(i\right),{\bf x'}}
-2\delta_{{\bf x},{\bf x'}}
+\delta_{{\bf x}-\hat{\sigma}\left(i\right),{\bf x'}}\right).
\label{threetwentyeight}
\end{equation}
{}From the equation we see that Poisson bracket is independent of time $x^0$:
\begin{equation}
\left\{\phi\left(x^0,{\bf x}\right),\pi\left(x^0,{\bf x'}\right)
\right\}_{\rm P.B.}=\delta_{{\bf x},{\bf x'}},
\label{threetwentynine}
\end{equation}
This corresponds to that the equal-time
commutation relation in quantum theory
is independent of time.\\
	The difference equations ~(\ref{threetwentysixa}) and
{}~(\ref{threetwentysixb}) conserve the following quantity:
\begin{eqnarray}
\cal{H}& = & \sum_{{\bf x}}{1 \over 2}
\pi^2\left(x^0,{\bf x}\right)+{\tau \over 2}\sum_{{\bf x},{\bf x'}}
\phi\left(x^0,{\bf x}\right){\bf \Omega}_{{\bf x},{\bf x'}}\;\;
\pi \left(x^0,{\bf x'}\right)\nonumber\\
&  &+{1 \over 2}\sum_{{\bf x},{\bf x'}}\phi\left(x^0,{\bf x }\right)
    {\bf \Omega}_{{\bf x},{\bf x'}}\;\;\phi\left(x^0,{\bf x'}\right)
+{g \over 4}
    \sum_{{\bf x}}\left\{\phi\left(x^0,{\bf x}\right)+\tau
    \pi\left(x^0,{\bf x}\right)\right\}^2\phi^2\left(x^0,{\bf x}\right)
    \nonumber\\
& = &\sum_{{\bf x}}{1 \over 2}
    \biggr[\pi^2\left(x^0,{\bf x}\right)+
       \sum^{3}_{i=1}\Delta^{R}_{i}\phi\left(x^0,{\bf x}\right)
       \left\{\Delta^{R}_{i}\phi\left(x^0,{\bf x}\right)+\tau\Delta^{R}_{i}
       \pi\left(x^0,{\bf x}\right)\right\}\nonumber\\
& &\;\;\;\;\;\;\;\;\;\;\;\;\;\;\;
   +m^2\phi\left(x^0,{\bf x}\right)
       \left\{\phi\left(x^0,{\bf x}\right)
          +\tau\pi\left(x^0,{\bf x}\right)
       \right\}
    \biggr]\nonumber\\
& &+{g \over 4}\sum_{{\bf x}}\left\{\phi\left(x^0,{\bf x}\right)+
    \tau\pi\left(x^0,{\bf x}\right)\right\}^2\phi^2\left(x^0,{\bf x}\right)
    \;.
\label{threethirty}
\end{eqnarray}

As we have seen above, the field with the interaction
{}~(\ref{threetwentyfour}) is formulated canonically and the conserved quantity
is obtained. However, we cannot find the time developing operator.

\noindent
	b) Next we assume
\begin{equation}
G\left(\phi\right)
=-{V   \left(x^0+\tau,{\bf x}\right)-V   \left(x^0-\tau,{\bf x}\right) \over
   \phi\left(x^0+\tau,{\bf x}\right)-\phi\left(x^0-\tau,{\bf x}\right)},
\label{threethirtyone}
\end{equation}
that satisfied Eq.~(\ref{threesix}), where we used the abbreviation:
$V\left(x^0,{\bf x}\right)=V\left(\phi\left(x^0,{\bf x}\right)\right)$.
In this case we cannot find the momentum conjugate to $\phi\left(x\right)$, so
as to rewrite the equation in a canonical form.  However, we can define
energy-momentum tensor $T_{\mu\nu}$, that satisfies the continuity condition:
\begin{equation}
\sum^{3}_{\mu,\rho=0}\eta^{\mu\rho}\Delta^{R}_{\mu}T_{\nu\rho}=0,
\label{threethirtytwo}
\end{equation}
\begin{equation}
T_{\rho\mu}=I_{\rho\mu}-\eta_{\rho\mu}K\left(\rho\right),
\label{threethirtythree}
\end{equation}
\begin{equation}
I_{\rho\mu}={1 \over 2}\left[\Delta^{R}_{\rho}\phi\left(x\right)
\Delta^{L}_{\mu}\phi\left(x\right)+\sigma\left(\mu\right)
\Delta^{R}_{\rho}\phi\left(x\right)\Delta^{R}_{\rho}
\Delta^{L}_{\mu}\phi\left(x\right)\right],
\label{threethirtyfour}
\end{equation}
\begin{eqnarray}
K\left(\rho\right)& = &{1 \over 2}\sum^{3}_{\mu,\nu=0}\eta^{\mu\nu}
\Delta^{R}_{\mu}\phi\left(x\right)
\Delta^{R}_{\nu}\phi\left(x\right)
+{1 \over 2}\sigma\left(\rho\right)\Delta^{L}_{\rho}\phi\left(x\right)
G\left(\phi\left( x \right)\right)\nonumber\\
& & +{1 \over 2}\sum^{x^{\rho}-\sigma\left(\rho\right)}_{y^{\rho}=-\infty}
\left(\Delta^{R}_{\rho}\phi\left(y\right)+
\Delta^{L}_{\rho}\phi\left(y\right)\right)
G\left(\phi\left(y\right)\right){\biggr \vert}
_{y^{\sigma}=x^{\sigma}\left(\rho \neq \sigma\right)}.
\label{threethirtyfive}
\end{eqnarray}

Therefore, the conserved quantities are
\begin{eqnarray}
{\cal P}_{0}& = &\sum_{{\bf x}}T_{00}\nonumber\\
& = &{1 \over 2}\sum_{{\bf x}}\biggr\{\left(
      \Delta^{R}_{0}\phi\left(x\right)\right)^2+\sum^{3}_{i=1}
      \Delta^{R}_{i}\phi\left(x^0+\tau,{\bf x}\right)
      \Delta^{R}_{i}\phi\left(x^0,{\bf x}\right)\nonumber\\
& &\;\;\;\;\;\;\;\;\;\;+V\left(x^0,{\bf x}\right)
                       +V\left(x^0+\tau,{\bf x}\right)\biggr\},
\label{threethirtysix}
\end{eqnarray}
\begin{eqnarray}
{\cal P}_{k}& = &\sum_{{\bf x}}T_{k0}\nonumber\\
            & = &-{1 \over 2\sigma\tau}\sum_{{\bf x}}
\left\{\phi\left(x^0,{\bf x}+\hat{\sigma}\left(k\right)\right)-
      \phi\left(x^0,{\bf x}-\hat{\sigma}\left(k\right)\right)\right\}
      \phi\left(x^0-\tau,{\bf x}\right).
\label{threethirtyseven}
\end{eqnarray}
These quantities correspond to energy and momentum in the continuous limit
respectively.

We have found the conserved quantities in the case of arbitrary
interaction $V\left(\phi\right)$.
However, we could not find the canonical formalism
and the time developing operator
\subsection{{\bf Quantum field theory with interaction}}\label{perturb}
	In this section we give up for a moment to obtain conserved
quantities, and try to formulate the quantum field theory with interaction
on discrete spacetime.
For this purpose it is necessary to have the time developing operator.  We
assume the existence of Lagrangian, i.e.,
\begin{equation}
G\left(\phi\right)=-{\partial V \over \partial \phi}.
\label{threethirtyeight}
\end{equation}
In a very simple case
\begin{mathletters}
\begin{equation}
G\left(\phi\right)=-g\phi^3-m^2\phi
\label{threethirtyninea}
\end{equation}
or
\begin{equation}
V\left(\phi\right)={g \over 4}\phi^4+{m^2 \over 2}\phi^2,
\label{threethirtynineb}
\end{equation}
\end{mathletters}
it is supposed from the case $V\left(x\right)={\Lambda}x^4/4$ in
Sec.\ II that in high "energy" region the system has
no conserved quantity and behaves chaotically.

Now we assume the time developing operator $U$ as follows:
\begin{eqnarray}
U\left(\phi,\pi\right)& = &
     \exp{\left\{-\tau\sum_{{\bf x}}{1 \over 2}\pi^2\left(x^0,{\bf x}\right)
                \;\;, \;\;\;\;\;
          \right\}_{\rm P.B.}
         }\;\;\;\nonumber\\
& \times &
     \exp{\left\{-\tau\sum_{{\bf x}}\left[{1 \over 2}\sum^{3}_{i=1}
          \left(\Delta^{R}_{i}\phi\left(x^0,{\bf x}\right)\right)^2
             +V\left(x^0,{\bf x}\right)\right]\;\;,
             \;\;\;\;\;
          \right\}_{\rm P.B.}
     }\;\;\;,
\label{threefourty}
\end{eqnarray}
where
\begin{equation}
\pi\left(x\right)=\Delta^{R}_{0}\phi\left(x\right),
\label{threefourtyone}
\end{equation}
and Poisson bracket is the same that of free field:
\begin{equation}
\left\{\phi\left(x^0,{\bf x}\right),\pi\left(x^0,{\bf x'}\right)
\right\}_{\rm P.B.}
=\delta_{{\bf x},{\bf x'}}.
\label{threefourtytwo}
\end{equation}
Using the operator we see that the field equation~(\ref{threeone}) or
{}~(\ref{threefive}) is equivalent to the following canonical equations:
\begin{mathletters}
\begin{equation}
\phi\left(x^0+\tau,{\bf x}\right)=U\phi\left(x^0,{\bf x}\right)=
\phi\left(x^0,{\bf x}\right)+\tau\pi\left(x^0,{\bf x}\right),
\label{threefourtythreea}
\end{equation}
\begin{eqnarray}
\pi\left(x^0+\tau,{\bf x}\right)& = & U\pi\left(x^0,{\bf x}\right)\nonumber\\
& = & \pi\left(x^0,{\bf x}\right)+\tau\biggr[\sum^{3}_{i=1}\Delta^{R}_{i}
      \Delta^{L}_{i}\phi\left(x^0+\tau,{\bf x}\right)
      +G\left(x^0+\tau,{\bf x}\right)\biggr].
\label{threefourtythreeb}
\end{eqnarray}
\end{mathletters}
Needless to say, Eq.~(\ref{threefourtythreea}) is the same as
Eq.~(\ref{threefourtyone}).

If the time developing operator is written in the following form:
\begin{equation}
U^{-1}=e^{-\tau \left\{H,\;\;\;\;\;\right\}_{\rm P.B.}}.
\label{threefourtyfour}
\end{equation}
$H$ is conserved and interpreted as the energy or time developing generator.
As is mentioned above, this quantity $H$ is not supposed to exist always.

The quantization of the field is straightforward.  The canonical
commutation relation is
\begin{equation}
\left[\phi\left(x^0,{\bf x}\right),\pi\left(x^0,{\bf x'}\right)\right]
=i\delta_{{\bf x},{\bf x'}},
\label{threefourtyfive}
\end{equation}
and the time developing operator is
\begin{eqnarray}
U\left(\phi,\pi\right)
&=&\exp{\left[-i\tau\sum_{{\bf x}}\left\{{1 \over 2}
             \sum^{3}_{i=1}\left(\Delta^{R}_{i}
             \phi\left(x^0,{\bf x}\right)\right)^2
             +V\left(x^0,{\bf x}\right)\right\}
          \right]
         }\nonumber\\
& &\;\;\;\;\;\times \exp{\left[-i\tau\sum_{{\bf x}}{1 \over 2}
                      \pi\left(x^0,{\bf x}\right)^2
                     \right]
                     },
\label{threefourtysix}
\end{eqnarray}
\begin{mathletters}
\begin{equation}
\phi\left(x^0+\tau,{\bf x}\right)=U\left(\phi,\pi\right)^{-1}
\phi\left(x^0,{\bf x}\right)U\left(\phi,\pi\right),
\label{threefourtysevena}
\end{equation}
\begin{equation}
\pi\left(x^0+\tau,{\bf x}\right)=U\left(\phi,\pi\right)^{-1}
\pi\left(x^0,{\bf x}\right)U\left(\phi,\pi\right).
\label{threefourtysevenb}
\end{equation}
\end{mathletters}
These equations are the same as Heisenberg equations~(\ref{threefourtythreea})
and (\ref{threefourtythreeb}).

After we have obtained the time developing operator, the next problem
is how to calculate the scattering matrix.  Following the method used in the
case of continuous spacetime, we express the field in the interaction
representation.
A physical state in Schr\"{o}dinger representation
at time $x^0$ is translated to the state at
the next time $x^0+\tau$ by the time developing
operator in Schr\"{o}dinger representation:
\begin{equation}
\mid\psi_{S},x^0+\tau\rangle=U\left(\phi_{S},\pi_{S}\right)
\mid\psi_{S},x^0\rangle .
\label{threefifty}
\end{equation}
$U\left(\phi_{S},\pi_{S}\right)$ is given by Eq.~(\ref{threefourtysix})
in which $\phi$ and $\pi$ of Heisenberg operator are replaced by $\phi_{S}$
and $\pi_{S}$ of Schr\"{o}dinger operator respectively.  We simply assume
that $\phi$ and $\pi$ coincide with $\phi_{S}$ and $\pi_{S}$ at
$x^{0}=0$.

Now the time developing operator in Schr\"{o}dinger representation
is rewritten as follows:
\begin{eqnarray}
U^{-1}\left(\phi_{S},\pi_{S}\right)& = &
\exp{\left[i\tau\sum_{{\bf x}}{1 \over 2}\pi_{S}^{2}\left({\bf x}\right)
     \right]
    }
\exp{\left[i\tau\sum_{{\bf x}}\left\{{1 \over 2}\sum^{3}_{i=1}
                                 \left(\Delta^{R}_{i}\phi_{S}\left({\bf x}
                                                           \right)
                                 \right)^2
                                 +V\left(\phi_{S}\left({\bf x}\right)\right)
                              \right\}
     \right]
    }\nonumber\\
& = &\exp{\left[i\tau\sum_{{\bf x}}{1 \over 2}\pi^{2}_{S}\left({\bf x}\right)
          \right]
         }
     \exp{\left[i\tau\sum_{{\bf x}}\left\{{1 \over 2}\sum^{3}_{i=1}
                                 \left(\Delta^{R}_{i}\phi_{S}\left({\bf x}
                                                                  \right)
                                      \right)^2+{1 \over 2}m^2\phi^{2}_{S}
                                      \left({\bf x}\right)
                                   \right\}
          \right]
       }\nonumber\\
& &\;\;\;\times\exp{\left[i\tau\sum_{{\bf x}}V_{I}\left(\phi_{S}\right)
          \right]
       }\nonumber\\
& = &\exp{\left(i{\tau}H_{0}\right)}
     \exp{\left[i\tau\sum_{{\bf x}}V_{I}\left(\phi_{S}\right)
          \right]
         }\nonumber\\
& = &U_{0}^{-1}\left(\phi_{S},\pi_{S}\right)
     \exp{\left[i\tau\sum_{{\bf x}}V_{I}\left(\phi_{S}\right)
          \right]
        },
\label{threefiftyone}
\end{eqnarray}
where
\begin{equation}
V_{I}\left(\phi_{S}\right)=V\left(\phi_{S}\right)-{1 \over 2}m^2\phi^{2}_{S}.
\label{threefiftytwo}
\end{equation}
and $U_{0}\left(\phi,\pi\right)$ is defined in Eq.~(\ref{threetwentyone})
.  The field operators in interaction representation are obtained from those
in Schr\"{o}dinger representation using the time developing operator of
free field in Schr\"{o}dinger representation:
\begin{mathletters}
\begin{equation}
\phi_{I}\left(x^0,{\bf x}\right)=U^{-x^0 / \tau}_{0}
\left(\phi_{S},\pi_{S}\right)
\phi_{S}\left({\bf x}\right)U^{x^0 / \tau}_{0}
\left(\phi_{S},\pi_{S}\right),
\label{threefiftythreea}
\end{equation}
\begin{equation}
\pi_{I}\left(x^0,{\bf x}\right)=U^{-x^0 / \tau}_{0}
\left(\phi_{S},\pi_{S}\right)
\pi_{S}\left({\bf x}\right)U^{x^0 / \tau}_{0}
\left(\phi_{S},\pi_{S}\right).
\label{threefiftythreea}
\end{equation}
Therefore, we see that the time developing operator for the field operator
in interaction representation is the same as that for free field in
Schr\"{o}dinger representation:
\begin{mathletters}
\end{mathletters}
\begin{equation}
\phi_{I}\left(x^0+\tau,{\bf x}\right) =
U_{0}^{-1}
\left(\phi_{I},\pi_{I}\right)
\phi_{I}\left(x^0,{\bf x}\right)
U_{0}
\left(\phi_{I},\pi_{I}\right),
\label{threefiftyfoura}
\end{equation}
\begin{equation}
\pi_{I}\left(x^0+\tau,{\bf x}\right)=
U_{0}^{-1}
\left(\phi_{I},\pi_{I}\right)
\pi_{I}\left(x^0,{\bf x}\right)
U_{0}
\left(\phi_{I},\pi_{I}\right),
\label{threefiftyfourb}
\end{equation}
\end{mathletters}
where
\begin{eqnarray*}
U_{0}\left(\phi_{I},                        \pi_{I}\right) & \equiv &
U_{0}\left(\phi_{I}\left(x^0,{\bf x}\right),\pi_{I}\left(x^0,{\bf x}\right)
           \right)\nonumber\\
& = & U^{-x^0 / \tau}_{0}\left(\phi_{S}\left({\bf x}\right),
                               \pi _{S}\left({\bf x}\right)\right)
                    U_{0}\left(\phi_{S}\left({\bf x}\right),
                               \pi _{S}\left({\bf x}\right)\right)
      U^{ x^0 / \tau}_{0}\left(\phi_{S}\left({\bf x}\right),
                               \pi _{S}\left({\bf x}\right)\right).
\end{eqnarray*}

The physical state of interaction representation is expressed as
follows:
\begin{eqnarray}
\mid\psi_{I},x^0\rangle
& = & U_{0}^{-x^0/\tau}\left(\phi_{S},\pi_{S}\right)
      \mid\psi_{S},x^0{\rangle}\nonumber\\
& = & U_{0}^{-x^0/\tau}\left(\phi_{S},\pi_{S}\right)
      U    ^{ x^0/\tau} \left(\phi_{S},\pi_{S}\right)
      \mid\psi_{S},0{\rangle}\nonumber\\
& = &U_{0}^{-x^0/\tau}\left(\phi_{S},\pi_{S}\right)
     \left[\exp{\left\{-i\tau\sum_{{\bf x}}
                    V_{I}\left(\phi_{S}\left({\bf x}\right)\right)
                \right\}
               }
        U_{0}\left(\phi_{S},\pi_{S}\right)
     \right]^{x^0/\tau}
     \mid\psi_{S},0\rangle\;\;\nonumber\\
& = &\prod^{\left(x^0/\tau\right)-1}_{l=0}
     \exp{\left\{-i{\tau}\sum_{{\bf x}}V_{I}\left(\phi_{I}
                                        \left(x^0-l\tau,{\bf x}\right)
                                     \right)
          \right\}
         }\mid\psi_{S},0\rangle.
\label{threefiftyfive}
\end{eqnarray}
The physical state of interaction representation is thus developed in the
following form:
\begin{equation}
\mid \psi_{I},x^0+\tau \rangle=
     \exp{\left\{-i{\tau}\sum_{{\bf x}}V_{I}\left(\phi_{I}
                                              \left(x^0+\tau,{\bf x}\right)
                                           \right)
          \right\}
         }\mid \psi_{I},x^0 \rangle.
\label{threefiftysix}
\end{equation}

The scattering matrix element between an initial state and a final
state is written as
\begin{eqnarray}
S_{fi} & \equiv & \langle \psi^{final  }_{I},x^0=+\infty
                     \mid \psi^{initial}_{I},x^0=-\infty \rangle
\nonumber\\
& = &\langle \psi^{final}_{I},x^0=-\infty \mid
     \prod^{\infty}_{l=-\infty}e^{-i{\tau}\sum_{{\bf x}}
                                    V_{I}\left(\phi_{I}
                                            \left(l\tau,{\bf x}\right)
                                         \right)
                                 }
     \mid \psi^{initial}_{I},x^0=-\infty \rangle.
\label{threefiftyseven}
\end{eqnarray}
If $V_{I}$ contains a small parameter, the scattering matrix is expanded in
a power series of $V_{I}$ :
\begin{eqnarray}
S_{fi} & = &\langle \; f \mid i \; \rangle
         -  i{\tau} \langle \; f \mid \sum_{x}
            V_{I}\left(x\right)\mid i \; \rangle \nonumber\\
       & - &{{\tau}^2 \over 2}\langle \; f \mid
            \sum_{x,y}\left[\theta\left(x^0-y^0\right)V_{I}\left(x\right)
                                                      V_{I}\left(y\right)
                           +\theta\left(y^0-x^0\right)V_{I}\left(y\right)
                                                      V_{I}\left(x\right)
                      \right]
            \mid i \; \rangle \nonumber\\
       & + & \cdots.
\label{threefiftyeight}
\end{eqnarray}
where we wrote $\mid i \; \rangle$ and $\mid f \; \rangle$ for
$\mid \psi^{initial}_{I},x^0 = -\infty \rangle\;\;$ and
$\mid \psi^{final}  _{I},x^0 = -\infty \rangle$ for simplicity and used the
abbreviation :
\[V_{I}\left(x\right)=V\left(\phi_{I}\left(x^0,{\bf x}\right)\right).\]
The summation $\sum_{x}$ means the sum over $x^0$ and ${\bf x}$.

%
%
\section{{\bf Discussion}}
\label{sect4}

We formulated quantum mechanics on discrete time
in the framework of canonical formalism in Sec.\ \ref{sect2}.
We will briefly comment here on the relation between
the canonical method and path integral quantization
on discrete time.

The transition matrix element on usual continuous
time is given by the
following formula in path integral formulation,
\begin{equation}
   \langle x_{f},t_{f} | x_{i},t_{i} \rangle =
   \int_{(x_{f},t_{f}) \sim (x_{i},t_{i})}
   {\cal D}x \hspace{1mm} e^{iS},
  \label{path}
\end{equation}
where
\( S \)
is an action.
The quantization on discrete time by path integral is
obtained only by replacing the continuous action in
Eq.\ (\ref{path}) by the discrete action (\ref{action}).
The path integral quantization is equivalent to the
canonical one in continuum theories,
but we have to check the equivalence for discrete time.
This can be easily seen as follows.
The transition matrix element is written like
\begin{equation}
   \langle x_{f},t_{f} | x_{i},t_{i} \rangle =
   \langle x_{f} | \prod e^{-i\tau V(\hat{x})}
                         e^{-i\tau \hat{p}^{2}/2}
                 | x_{i} \rangle
   \label{ourcan}
\end{equation}
in our canonical formalism.
We insert complete sets
\( 1 = \int dx | x \rangle \langle x | \)
and
\( 1 = \int dp | p \rangle \langle p | \)
between exponentials and write them as an integral
over
c-numbers.
After the integrations over $p$'s
using
\( \langle x | p \rangle = e^{ixp}/\sqrt{2\pi} \),
we have (\ref{path}) with the discrete
action (\ref{action}).

In a continuous time theory, an evolution operator
\( U_{\rm cont}(t) \)
is given by
\begin{equation}
   U_{\rm cont}(t) = \exp(-it H_{\rm cont})
               = [\exp(-i\Delta tH_{\rm cont})]^{t/\Delta t},
\end{equation}
where \( H_{\rm cont} = \hat{p}^2/2 + V(\hat{x}) \).
We have to use the approximation :
\begin{equation}
  \exp(-i\Delta t H_{\rm cont})
       \cong \exp(-i\Delta t V(\hat{x}))
             \exp(-i\Delta t \hat{p}^{2}/2),
   \label{approx}
\end{equation}
for short time distance
\( \Delta t \) in order to show the equivalence.
The calculation for the proof is completely
same as above \cite{disc1}.
However, the right hand side of Eq.\ (\ref{approx}) is an exact time
evolution operator in our framework :
\begin{equation}
   \exp(-i\tau H_{\tau}) =
           \exp(-i\tau V(\hat{x}))
           \exp(-i\tau \hat{p}^{2}/2),
\end{equation}
though \( H_{\tau} \) becomes very complicated
(see Eq.\ (\ref{htau_exp})).
If we consider
\( \exp(-i\tau H_{\rm cont}) \) with
discrete variables as an time evolution operator,
the equation of motion becomes very complicated.

Generally, it is not evident what kind of quantity
corresponds to usual momentum in a discrete time theory.
We see that the usual discrete approximation holds
exactly under the definition of momentum (\ref{reasonable_p})
for our simple equation of motion.
This definition is natural in this sense.
In other words,
this discretization scheme has a natural
continuum limit and is appropriate on discrete time.

In Sec.\ \ref{sect3}, we formulated quantum field theory
on discrete Minkowski spacetime.
Lattice theories are usually formulated on Euclidean
spacetime because of difficulties of indefinite metric
on Minkowski spacetime.
The relation between continuum theories on Euclidean and Minkowski spacetime
has been investigated by many authors \cite{disc2}-\cite{disc4}.
Sufficient conditions for
field operators of two spacetimes to be connected
with each other by simple analytic continuation
is given by Osterwalder and Schrader \cite{disc2}.
In their proof, Minkowski and Euclidean invariances
are essential.
We have no such invariances on discrete
spacetime and this simple relation does not hold.
As an example, the Euclidean propagator of free
scalar field is
\begin{eqnarray}
\lefteqn{
\langle0\mid T\left(
\phi\left(x\right)\phi\left(x'\right)
\right)\mid0\rangle
}
\;\;\;\;\;\;    \nonumber\\
& = & -
\left(\prod^{3}_{\mu=0}
     \int^{{  \pi \over \sigma\left(\mu\right)}}
          _{-{\pi \over \sigma\left(\mu\right)}}
     {\sigma\left(\mu\right) \over 2\pi}dk_\mu
\right)
\exp{\left\{i\sum_{\mu,\nu=0}^3 \delta^{\mu\nu}
                  k_\mu\left(x_\nu-x'_\nu\right)\right\}}\nonumber\\
& &  \times\left[\sum^{3}_{\mu,\nu=0}\delta^{\mu\nu}
                 {4 \over \sigma\left(\mu\right)\sigma\left(\nu\right)
                 }
             \sin{{\sigma\left(\mu\right)k_{\mu}} \over 2}
             \sin{{\sigma\left(\nu\right)k_{\nu}} \over 2}
             +m^2
         \right]^{-1}
    ,
\end{eqnarray}
and the Minkowski propagator is
\begin{eqnarray}
\lefteqn{
\langle0\mid T\left(
\phi\left(x\right)\phi\left(x'\right)
\right)\mid0\rangle
}
\;\;\;\;\;\;    \nonumber\\
& = & i
\left(\prod^{3}_{\mu=0}
     \int^{{  \pi \over \sigma\left(\mu\right)}}
          _{-{\pi \over \sigma\left(\mu\right)}}
     {\sigma\left(\mu\right) \over 2\pi}dk_\mu
\right)
\exp{\left\{-ik_\nu\left(x^\nu-x'^\nu\right)\right\}}\nonumber\\
& &  \times\left[\sum^{3}_{\mu,\nu=0}\eta^{\mu\nu}
                 {4 \over \sigma\left(\mu\right)\sigma\left(\nu\right)
                 }
             \sin{{\sigma\left(\mu\right)k_{\mu}} \over 2}
             \sin{{\sigma\left(\nu\right)k_{\nu}} \over 2}
             -m^2+i\epsilon
         \right]^{-1}
    .
\end{eqnarray}
They are not connected by analytic continuation.
We always have to be careful about the above
when we extract
Minkowski information from quantities calculated
on Euclidean spacetime.

%
%

\figure{
  Behavior of the iterative map of Eq.\ (2.33a,b)
with the potential $V(x)=\Lambda x^4/4$.
Variables are scaled so that $\tilde x=\tau\sqrt{\Lambda}\,x$
and $\tilde p=\tau^2\sqrt{\Lambda}\,p$.
Initial values are $\tilde p(0)=0$ and $\tilde
x(0)=0.17,\ 0.47,\ 0.62,\ 0.67,\ 0.76,\ 0.775,\ 0.80$.
For small initial values of $\tilde x(0)$, the map
draws a smooth curve, which indicates the existence
of a conserved quantity $H_\tau$.
The chaotic behavior for larger initial values of
$\tilde x(0)$ suggests the divergence of the series of $H_\tau$.
  \label{fig1}
}

\end{document}